\documentclass[letterpaper]{article} 
\usepackage{aaai23}  
\usepackage{times}  
\usepackage{helvet}  
\usepackage{courier}  
\usepackage[hyphens]{url}  
\usepackage{graphicx} 
\usepackage{natbib}  
\usepackage{caption} 
\usepackage{algorithm}
\usepackage{algorithmic}

\usepackage{newfloat}
\usepackage{listings}
\DeclareCaptionStyle{ruled}{labelfont=normalfont,labelsep=colon,strut=off} 
\floatstyle{ruled}
\newfloat{listing}{tb}{lst}{}
\floatname{listing}{Listing}
\usepackage[russian,english]{babel}
\newcommand{\RUS}{\foreignlanguage{russian}}
\title{Tweets in Time of Conflict: A Public Dataset Tracking the Twitter Discourse on the War Between Ukraine and Russia}
\author{
    Emily Chen\textsuperscript{\rm 1,\rm 2},
    Emilio Ferrara\textsuperscript{\rm 1,\rm 2,\rm 3}
}
\affiliations{
    \textsuperscript{\rm 1} Department of Computer Science, University of Southern California\\
    \textsuperscript{\rm 2} Information Sciences Institute, University of Southern California\\
    \textsuperscript{\rm 3} Annenberg School of Communication, University of Southern California \\
    echen920@usc.edu, emiliofe@usc.edu

}

\begin{document}

\maketitle

\begin{abstract}
On February 24, 2022, Russia invaded Ukraine. In the days that followed, reports kept flooding in from laymen to news anchors of a conflict quickly escalating into war. Russia faced immediate backlash and condemnation from the world at large. While the war continues to contribute to an ongoing humanitarian and refugee crisis in Ukraine, a second battlefield has emerged in the online space, both in the use of social media to garner support for both sides of the conflict and also in the context of information warfare. In this paper, we present a collection of nearly half a billion tweets, from February 22, 2022, through January 8, 2023, that we are publishing for the wider research community to use. This dataset can be found at \textbf{\url{https://github.com/echen102/ukraine-russia}}. Our preliminary analysis on a subset of our dataset already shows evidence of public engagement with Russian state-sponsored media and other domains that are known to push unreliable information towards the beginning of the war; the former saw a spike in activity on the day of the Russian invasion, while the other saw spikes in engagement within the first month of the war. Our hope is that this public dataset can help the research community to further understand the ever-evolving role that social media plays in information dissemination, influence campaigns, grassroots mobilization, and much more, during a time of conflict.
\end{abstract}

\section{Introduction}
\subsection{Timeline of Key Events}
The tensions between Ukraine and Russia have been on the rise for decades, but have substantially increased in the past few years. Following the Soviet Union’s dissolution in 1991, Ukraine declared itself an independent country on August 24, 1991~\cite{sullivan2022russia}. 
In 2010, the pro-Russian presidential candidate, Viktor Yanukovich, was elected into office amid accusations of election fraud~\cite{reuters2022timeline,kahn2022viktor}. 
Yanukovich was ousted in 2014 after backing out of an agreement with the European Union (EU), an agreement which would have brought Ukraine closer to becoming a member of the EU, in favor of engaging with Russia~\cite{kahn2022viktor}. 

Shortly after Yanukovich was removed from office, Russia annexed Crimea from Ukraine against the wishes of the West, further exacerbating tensions between Russia and Ukraine. 
Russia claimed that this action was supported by the vast majority of the Crimean people, but this was also denounced as a fraudulent vote by Western countries and Ukrainian leadership~\cite{clinch2022russia,myers2014putin,collett-white_2014crimeans}.

Pro-Russian separatist groups in Donetsk and Luhansk in the Donbas region declared their independence, sparking the beginning of the war in Donbas~\cite{reuters2022timeline}. 
Several iterations of the Minsk agreement were signed between Ukraine and Russia in September 2014 in efforts to reach a resolution for the war; these efforts, however, have proven to be unsuccessful to this day~\cite{sullivan2022russia}. 

In 2019, the current President of Ukraine, Volodymyr Zelenskyy was elected to office, and one of his intentions was to end the conflict in Donbas~\cite{pereira2022what}. 

In late 2021 through early 2022, it became clear that Russia was amassing its forces near the Russian-Ukrainian borders. On February 21, 2022, Russia officially recognized the independence of Donetsk (Donetsk People’s Republic) and Luhansk (Luhansk People’s Republic)~\cite{westfall2022donetsk,bloomberg2022visual,sullivan2022russia}.

On February 24, 2022, Russia invaded Ukraine, an act that drew swift condemnation from many world leaders, including the EU and NATO allies~\cite{mcgee2022world}. 

The Russian-Ukrainian war has caused and continues to be, as of these days, a humanitarian emergency for Ukrainians in the country, and has also created a refugee crisis as Ukrainians turned refugees flee to neighboring countries in large numbers~\cite{schwirtz2022humanitarian,bloomberg2022visual,cengel2022history}. 

On March 2, 2022, Russia took control of Kherson, marking the first major Ukrainian city to fall to Russian troops~\cite{schwirtz2022first}. 

As a result of the Russian invasion, many Western powers have imposed sanctions on Russia in an attempt to deter and reverse Russian aggression, and Western companies have begun to withdraw their operations from Russia~\cite{funakoshi2022tracking}. 

As of this writing (April 2023), the war has continued to rage on, with causalities on both sides of the conflict~\cite{bigg_2023}.

\subsection{Social Media Activity}
During this time of conflict, information warfare and campaigns continued, both in the lead-up and in the aftermath of the invasion. 
This has taken place on a variety of social media platforms, including Twitter. 
Russian disinformation campaigns have been rampant on social media \cite{badawy2018analyzing, broniatowski2018weaponized, badawy2019characterizing, badawy2019falls, dutt2019senator, luceri2020detecting, ferrara2020characterizing, ezzeddine2022characterizing}. Evidence suggests they are being carried out both domestically and abroad~\cite{sharma2021identifying, scott2022war, la2023retrieving, pierri2022propaganda}.
However, Ukrainians have also waged their social media fight against Russia, and Russian President Vladimir Putin, by using social media platforms to promote the Ukrainian cause and garner international attention and support towards their current plight~\cite{cohen2022surge,garne2022war}.
Social media platforms have also taken actions to combat misinformation and disinformation in the wake of the conflict~\cite{cohen2022surge, bushwick2022russia, ferrara2022twitter, pierri2022does}. 

As the war continues to unfold, we hope that the research community can leverage our dataset to continue to combat misinformation, identify vulnerable communities and understand how the pervasiveness of social media has changed how modern conflict plays out both online and on the ground. These are just a few of many potential research directions that this data can help to answer, particularly in times when warfare is no longer limited by geographic bounds.
In this paper, we document how to access this dataset and provide an overview of basic statistics and general findings from our Twitter data collection. 

\section{Data Collection}

Our real-time Twitter data collection began on February 22, 2022. We used Twitter's streaming API v1.1\footnote{\url{https://developer.twitter.com/en/docs/twitter-api}} to track keywords of interest that were both trending and related to the conflict at the time of collection. This version of their API has since been deprecated. This list of keywords primarily focused on terms related to events and locations prevalent in the conflict within the first several months of the war; however, we updated our list as needed. Our list of keywords is not exhaustive, but we did our best to monitor ongoing discourse and consulted knowledgeable colleagues. Prior work has shown that the streaming API is not completely random, which results in some biases depending on collection and collection volume~\cite{mehrabi2021survey}. However, we still leveraged the Twitter streaming API endpoint as it is the most sustainable method of gathering data that still gives us an understanding of current Twitter discourse. 

We also leveraged Twitter's search API to collect tweets prior to February 22, 2022; while we will continue to collect historical tweets, due to Twitter's rate limit on the Academic Track search API, we are only able to use the search endpoint to collect 10 million tweets each month.\footnote{\url{https://developer.twitter.com/en/products/twitter-api/academic-research}} Our current data collection consists of over 570 million tweets, with more than 500 GB of raw data thus far. 
Our data collection concluded in late March 2023 due to recent modifications to the Twitter API. However, we remain optimistic that future API changes may enable us to recommence our data collection efforts in due course.

We host the dataset on our public GitHub repository so that other researchers are able to access tweet IDs that are pertinent to the Ukraine-Russia conflict, as misinformation and influence campaigns have already been detected in the narratives being pushed -- both on and off Twitter. To remain in compliance with Twitter's Terms \& Conditions, we are not permitted to publicly release text or metadata pertaining to any specific tweet outside of their tweet IDs. As a result, we have published tweet IDs so that researchers can use the Twitter API or third-party tools, such as \textit{Hydrator}\footnote{\url{https://github.com/DocNow/hydrator}} or \textit{Twarc},\footnote{\url{https://github.com/DocNow/twarc}} to obtain the raw tweet payload. We note here that for any tweet that has been removed from Twitter \cite{pierri2022does}, or if the account that posted a tweet has been deleted, suspended, or banned, researchers will be unable to retrieve the tweet content through Twitter's API. Researchers can also ensure that their datasets are in compliance with Twitter's Terms \& Conditions through Twitter's batch compliance endpoint to remove any tweets that have since been removed or hidden from public Twitter feeds.\footnote{\url{https://developer.twitter.com/en/docs/twitter-api/compliance/batch-compliance/quick-start}}

\subsection{Tracked Keywords}
We leveraged Twitter's trending topics and hashtags to identify and curate a list of keywords that are pertinent to the developing war between Ukraine and Russia. We then used this list to query Twitter's streaming API for any tweets that contained the keywords of interest in the tweet's text. Twitter's streaming API is capitalization agnostic, so we only need to track one capitalization permutation. We do, however, include some capitalization variations of keywords, particularly those that are not in English, to err on the side of caution. The list of the keywords that we tracked can be found in Table~\ref{keyword_table}, but as events continued to develop and unfold, we expanded this list while consulting those more intimately familiar with the conflict. The most up-to-date keyword list can be found on our GitHub repository. 

\begin{table}[t!]
    \centering
    \footnotesize
    \begin{tabular}{c|c|c}
    \textbf{Keywords} & \textbf{Tracked Since} & \textbf{Notes}\\
    \hline
    ukraine         & 2/22/22   &  \\
    russia          & 2/22/22   &  \\
    putin           & 2/22/22   &  \\
    soviet          & 2/22/22   &  \\
    kremlin         & 2/22/22   &  \\
    minsk           & 2/22/22   &  \\
    ukrainian       & 2/22/22   &  \\
    NATO            & 2/22/22   &  \\
    luhansk         & 2/22/22   &  \\
    donetsk         & 2/22/22   &  \\
    kyiv            & 2/22/22   &  \\
    kiev            & 2/22/22   &  \\
    moscow          & 2/22/22   &  \\
    zelensky        & 2/22/22   &  \\
    fsb             & 2/22/22   &  \\
    KGB             & 2/22/22   &  \\
    \RUS{Україна}   & 2/22/22   & Ukraine (uk)   \\
    \RUS{Киев}      & 2/22/22   & Kiev (ru)      \\
    \RUS{ФСБ}       & 2/22/22   & FSB (ru)        \\
    \RUS{Россия}    & 2/22/22   & Russia (ru)   \\
    \RUS{КГБ}       & 2/22/22   & KGB \\ 
    \RUS{Київ}      & 3/1/22    & Kiev (uk)   \\
    \RUS{україни}   & 3/1/22    & ukraine (uk) \\
    \RUS{Росія}     & 3/1/22    & Russia (uk) \\
    \RUS{кгб}       & 3/1/22    & kgb (uk/ru) \\
    \RUS{фсб}       & 3/1/22    & fsb (ru) \\
    SlavaUkraini    & 3/1/22    & Glory to Ukraine (ru) \\
    ukraini         & 3/1/22    & ukraine   \\
    U+1F1FA, U+1F1E6 & 3/1/22    & Ukrainian Flag emoji \\
    \RUS{Украина}   & 3/1/22    & Ukraine (ru) \\
    \RUS{украины}   & 3/1/22    & ukraine (ru) \\
    \RUS{Donbas}    & 3/1/22    & donbas \\
    \RUS{Донбас}    & 3/1/22    & donbas (uk) \\
    \end{tabular}
    \caption{Keywords that we actively tracked (v1.2 --- October 6, 2022). General notes and English translations provided in the notes section. Translations are followed with the original keyword language in parenthesis.}
    \label{keyword_table}
\end{table}

\section{Data \& Access Modalities}
\subsection{Release v1.2 (October 6, 2022)}

Our third release contains tweets from the inception of our data collection, February 22, 2022, 4 AM UTC, through the end of October 1, 2022. This consists of a total of \textbf{454,488,445} tweets, and we detail general statistics about this release in this section. We note that we have since released v1.5 of this dataset, but for the purposes of this paper, we focus on Release v1.2. The language breakdown of release v1.2 can be found in Table~\ref{tab:languages}, and the keywords that we were tracking are listed in Table~\ref{keyword_table}. Some of the keywords were added later to our live collection which impacted the language proportions that we observed in our dataset; we used Twitter's search API to obtain past tweets that we missed in our initial collection that also contain these keywords. We will be adding these additional tweet IDs in subsequent releases. 

\begin{table}[t!]
    \centering
    \begin{tabular}{c|c|c|c}
        \textbf{Language} & \textbf{ISO} & \textbf{No. tweets} & \textbf{\% total}\\
        \hline
        English & en & 321,088,619 & 70.65\%\\
        Spanish & es & 18,358,931 & 4.04\%\\
        French & fr & 17,857,397 & 3.93\%\\
        German & de & 14,533,854 & 3.2\%\\
        Italian & it & 11,589,565 & 2.55\%\\ 
        Undefined & (und) & 11,473,234 & 2.52\%\\
        Russian & ru & 9,968,421 & 2.19\%\\
        Japanese & ja & 9,113,466 & 2.01\%\\
        Ukrainian & uk & 8,016,384 & 1.76\%\\ 
        Turkish & tr & 6,219,988 & 1.37\%\\
        Portuguese & pt & 3,897,544 & 0.86\%\\
        Polish & pl & 3,411,167 & 0.75\%\\
        Dutch & nl & 1,837,698 & 0.4\%\\
        Indonesian & in & 1,607,514 & 0.35\%\\
        Chinese & zh & 1,430,735 & 0.31\%\\

    \end{tabular}
    \caption{ The top 15 languages in our dataset and the number of respective tweets. (v1.2 --- October 6, 2022)}
    \label{tab:languages}
\end{table}

\subsubsection{Language Distribution}
\begin{figure*}[t]
  \includegraphics[width=1.0\textwidth]{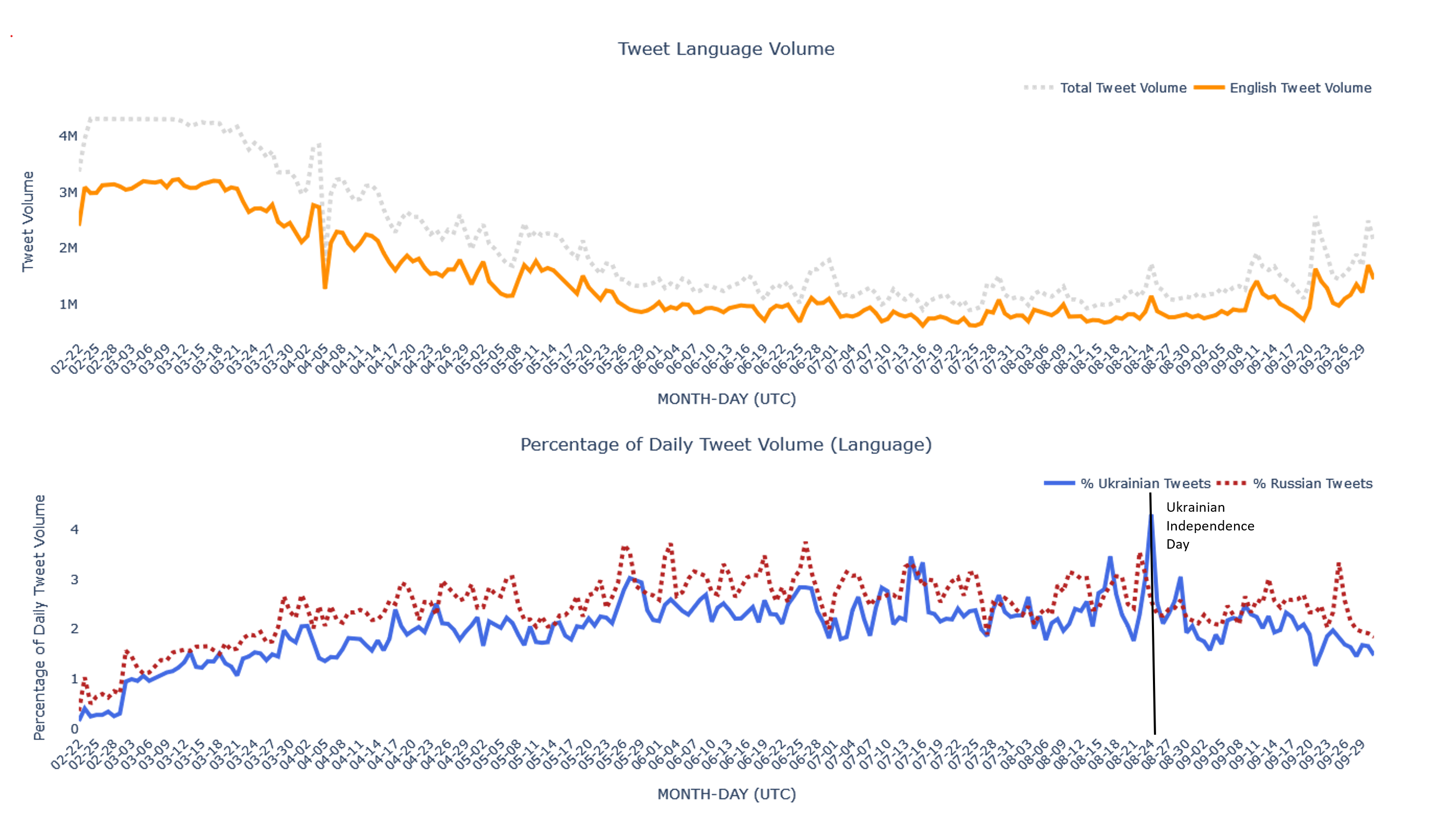}
\caption{Total volume of all tweets collected and volume of English, Ukrainian, and Russian tweets in our dataset. Note that we observe a decline in overall tweet volume collected per day around March 12, 2022; tweet volume was limited by Twitter prior to this date due to the high volume of tweets using the keywords of interest. Engagement with these keywords began to wane in the following months.}
\label{fig:lang}       
\end{figure*}

Twitter automatically attempts to tag each tweet with its language ISO and includes the found ISO in a tweet's metadata. When we investigate the language distribution of the tweets, we find that English is the predominant language that is identified. This aligns with our expectations, as most of the keywords that we were initially tracking were all in English. When we added keywords in the Ukrainian and Russian languages on March 1, 2022, we saw a significant increase in the percentage of Ukrainian and Russian tweets in our dataset. We notice that the volume of tweets we collect initially averages over 4 million tweets collected each day, which we attribute to Twitter limiting the number of tweets we can collect due to high tweet volume~\cite{osome2022}; however, we see a gradual decline in the volume of tweets we collect as the Ukraine-Russia war continued to rage on.

We also observe fluctuations in the volume of tweets posted in English, Ukrainian, and Russian, which, on an hourly granularity, matches the circadian patterns of the countries that predominantly speak that language and have a presence on Twitter. The English tweet volumes see an increase during general waking hours that follow United States time zones. This corroborates our finding, which we discuss later in this paper, that most tweets originate from the United States; Russian and Ukrainian tweets follow similar activity patterns, as Ukraine and Russia are in similar time zones. 

Our data also shows that spikes of tweets in a particular language generally occur alongside major real-world events. When we examine the percentage of tweets that are written in Ukrainian and Russian, in general, Russian tweets make up a slightly larger percentage of the total number of tweets we collected on any given day. However, there are several instances where Ukrainian tweets outnumber Russian tweets. For example, on August 24, 2022, Ukrainians celebrated their independence day, which correlates with the increased percentage of Ukrainian tweets on that day (\textit{cf.} Figure~\ref{fig:lang}, bottom)~\cite{hayda_2022,john_kesaieva_2022}. 

\subsubsection{Hashtags}
\begin{table}[t!]
    \centering
    \begin{tabular}{c}
        \textbf{Top 15 Hashtags}\\
        \hline
        ukraine \\ 
        russia \\
        putin \\
        standwithukraine \\
        ukrainerussiawar\\
        nato\\
        russian\\
        ukrainian \\
        kyiv\\
        ukrainewar\\
        zelensky\\
        mariupol\\
        stoprussia\\
        slavaukraini\\
        tigray\\

    \end{tabular}
    \caption{ The top 15 hashtags used in our dataset, are listed in decreasing order from top to bottom. (v1.2 --- October 6, 2022)}
    \label{tab:hashtags}
\end{table}

Table~\ref{tab:hashtags} lists the top 15 hashtags that we find used in our dataset. 
These hashtags are comprised of hashtags that are referring to groups involved or affiliated with the Ukraine-Russia war (e.g., \#ukraine, \#russia, \#nato), and others are in support of Ukraine and against the war (e.g., \#standwithukraine, \#stoprussia). Other commonly used hashtags refer to locations in Ukraine that have been at the center of conflict during the war (e.g., \#kyiv, \#mariupol). The hashtag \#tigray draws attention to the concurrent Tigray war and humanitarian crisis occurring in Ethiopia, with some wondering why the Ukrainian War has overshadowed the conflict in Ethiopia~\cite{cheng_anna_2022,walsh_2022}. %

Outside of these hashtags, other highly prevalent hashtags include \#anonymous, which referred to the announcement by the decentralized online hacking group Anonymous that they would be targeting Russian government websites and infrastructure while aiding Ukrainians~\cite{pitrelli2022anonymous}. 

\begin{figure*}[t]
  \includegraphics[width=1.0\textwidth]{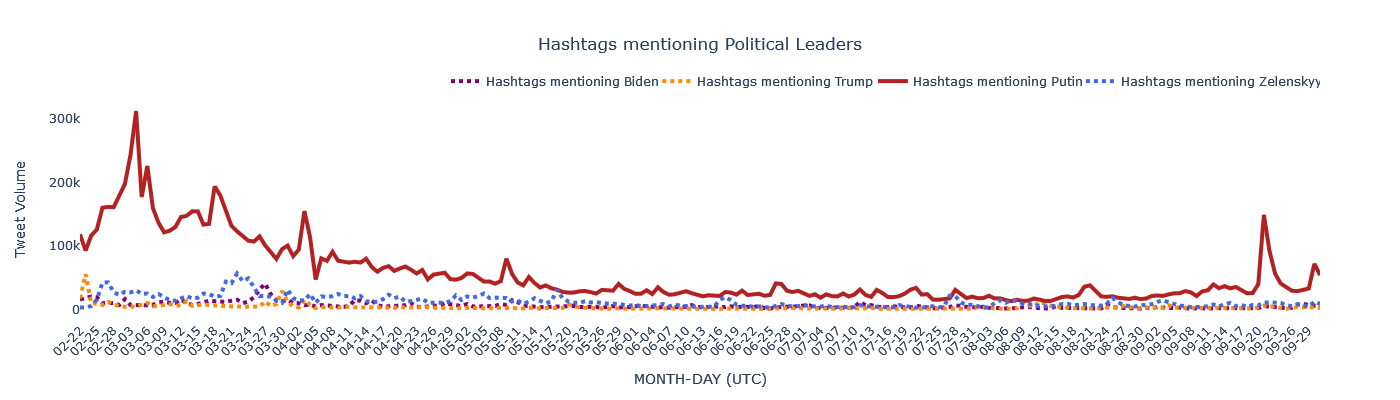}
\caption{Total volume of tweets that use the hashtags \#biden, \#trump, \#putin and \#zelenskyy. For \#zelenskyy, we also counted \#zelensky. The capitalization of a hashtag did not matter, as we lower-cased all hashtags.}
\label{fig:pol_figure_hts}       
\end{figure*}

When we compare the number of tweets using a hashtag containing the last name of the current political leaders of Russia and Ukraine, we can see that \#putin is much more widely used than \#zelenskyy. 
To further understand the magnitude of tweets that include \#putin, we also plotted the number of tweets that use \#biden for the current President of the United States, Joe Biden, and the number of tweets that use \#trump, referring to the former President of the United States, Donald Trump (see Figure~\ref{fig:pol_figure_hts}). 

In Figure~\ref{fig:pol_figure_hts}, we can see that \#putin is used the most frequently, and experiences volatile usage which, in general, reacts to breaking news and events. There is a spike of engagement with Putin-related hashtags on March 4, 2022, and a manual inspection of these tweets reveal that many users were using the hashtag \#SafeAirliftUkraine in tandem with \#StopPutin, requesting the President of the United States, Joe Biden, to help with the evacuation of Ukrainians. This coincides with reports on March 4 that, despite discussions on temporary ceasefires the days before to establish humanitarian corridors to aid civilian evacuations, Russia was not complying with the agreements that were made~\cite{stern2022limited,stern2022russia}. While hashtags using Biden and Trump see some use, the use of a hashtag mentioning Zelenskyy comes the closest to matching and sometimes slightly surpassing the number of tweets that use hashtags mentioning Putin. Given that Zelenskyy is the current president of Ukraine, these findings fall in line with our expectations. 

\subsubsection{Locations of Tweeters}
In each tweet's metadata that Twitter's API returns, we are given information about the user who posted a tweet. This will sometimes contain geolocation data from the user's post, but in previous works, we have observed less than 1\% of our dataset is returned with Twitter's own geolocation information~\cite{jiang2020political}. Thus, if a user manually provides a location in their user profile, we use this as a proxy to determine where a user is located~\cite{jiang2020political}. While this is not a perfect method, this is one of the only ways we can ascertain a user's general or affiliated physical location on a larger scale. 

\begin{table}[t!]
    \centering
    \begin{tabular}{ccc}
        \textbf{Retweeted Country} & \textbf{$\rightarrow$} & \textbf{Retweeter Country}\\
        \hline
        United States & $\rightarrow$ & United States \\
        United Kingdom & $\rightarrow$ &  United Kingdom \\
        Ukraine & $\rightarrow$ & United States \\
        United Kingdom &$\rightarrow$ &  United States \\
        France & $\rightarrow$ & France \\
        United States &$\rightarrow$ &  Canada \\
        United States & $\rightarrow$ & United Kingdom \\ 
        India &$\rightarrow$ &  India \\
        Ukraine & $\rightarrow$ & Ukraine \\
        Germany &$\rightarrow$ & Germany \\
        Italy & $\rightarrow$ & Italy \\
        Canada & $\rightarrow$ & United States \\
        Turkey & $\rightarrow$ & Turkey \\
        Ukraine & $\rightarrow$ & United Kingdom \\
        Spain & $\rightarrow$ & Spain \\

    \end{tabular}
    \caption{ The top 15 locations for retweets where a location for both the retweeted and retweeter were identifiable. Locations are listed in decreasing order from top to bottom. (v1.2 --- October 6, 2022)}
    \label{tab:location_rt}
\end{table}

\begin{table}[t!]
    \centering
    \begin{tabular}{ccc}
        \textbf{Quoted Country} & \textbf{$\rightarrow$} & \textbf{Quoter Country}\\
        \hline
        United States & $\rightarrow$ & United States \\
        United Kingdom & $\rightarrow$ &  United Kingdom \\
        United Kingdom &$\rightarrow$ &  United States \\
        France & $\rightarrow$ & France \\
        Ukraine & $\rightarrow$ & United States \\
        United States & $\rightarrow$ & United Kingdom \\ 
        United States &$\rightarrow$ &  Canada \\
        Germany & $\rightarrow$ & Germany \\
        India & $\rightarrow$ & India \\
        United States & $\rightarrow$ & Japan \\
        Japan & $\rightarrow$ & Japan \\
        Spain & $\rightarrow$ & Spain \\
        Russia & $\rightarrow$ & United States \\
        Canada & $\rightarrow$ & Canada \\
        Italy & $\rightarrow$ & Italy \\
    \end{tabular}
    \caption{ The top 15 locations for quoted tweets where a location for both the quoted user and user quoting a tweet were identifiable. Locations are listed in decreasing order from top to bottom. (v1.2 --- October 6, 2022)}
    \label{tab:location_qtd}
\end{table}
We find that most tweets originate from the United States and the United Kingdom, which are primarily English-speaking countries. Given that the vast majority of our initial keywords were in English, this matches our expectations. 
There are two different ways a user can retweet a post -- a user can simply retweet a post without adding additional commentary (we refer to this type of tweet as \textit{retweets}) or a user can add additional text to the post (which we refer to as a \textit{quoted} tweet). 
If user A retweets or quotes user B, we consider the direction of the retweet or quote to be from user B to user A. 
For quoted and retweeted tweets, we list the top 15 locations in Table~\ref{tab:location_rt} where we were able to identify a location for both the retweeted and the retweeter. We do the same for quoted tweets in Table~\ref{tab:location_qtd}. 

Interestingly, we find that users in the United States quote users from Ukraine 3.34 times more than they quote users from Russia; users that are retweeting tweets from the United States are 9.11 times more likely to retweet users from Ukraine as opposed to users from Russia. Users from the United States retweeting users from Russia is the 25th most frequent location pairing. 

\subsubsection{Domain Sharing}
\begin{figure*}[t]
  \includegraphics[width=1.0\textwidth]{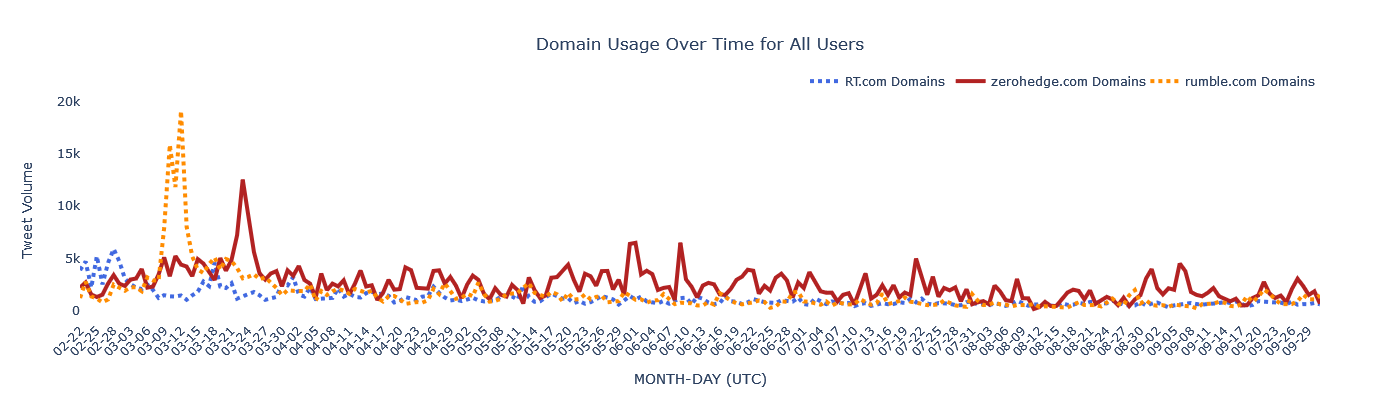}
\caption{Total volume of tweets that share specific domains in our dataset. RT.com is a known Russian state-owned media company, while MBFC has classified rumble.com as questionable and zerohedge.com as prone to conspiracy and pseudoscience.}
\label{fig:doms}       
\end{figure*}

Finally, we take a look at the domains that users post as a part of their tweets. We consider any domain that is included within a user's tweet as a domain that the user has shared, including domains that are part of a retweet. Within the top 50 domains that were shared, we find several domains that are known Russia state-sponsored media (www.RT.com), or tagged as questionable (www.rumble.com) and prone to conspiracies and pseudoscience (www.zerohedge.com) by Media Bias / Fact Check (MBFC)~\cite{mbfc}. MBFC is an independently operated website that classifies domains into certain factual and political lean categories, based on specific criteria that are described further on their per domain synopsis~\cite{mbfc}. 

While the number of tweets that interact with these particular domains make up a small proportion of all tweets posted about the Ukraine-Russia war, we do see a fairly consistent number of tweets that interact with these domains over time (\textit{cf}. Figure~\ref{fig:doms}). ZeroHedge is the most consistently posted over our observation period, but we do note that these domains experience spikes in engagement occurred within the first month of the war, with Rumble being shared in the most tweets in early March. 
\subsection{Access}
Our dataset can be found on GitHub at: \textbf{\url{https://github.com/echen102/ukraine-russia}}. 

Due to recent changes to the Twitter API, we are no longer able to update this dataset with new data as of late March 2023. We hope to resume our collection and regular updates in the future if subsequent API changes allow us to do so. Given that monthly rate limits have significantly decreased, we suggest interested parties take a random sample of tweet IDs to hydrate over the desired time frame. 

Researchers who would like to use our collected data must remain in compliance with Twitter's Terms \& Conditions,\footnote{\url{https://developer.twitter.com/en/developer-terms/agreement-and-policy}} and agree upon the terms of usage as outlined in the accompanying license. 

\subsection{Ethical Considerations}
As mentioned earlier in this paper, we are only able to publish the tweet IDs associated with our collected tweets as a part of our data collection. This is to ensure that we remain in compliance with Twitter's developer terms of service, which restrict us from publicly publishing any individual tweet information outside of the tweet's unique ID. All tweets that can be retrieved by the end user are tweets that are publicly accessible -- any tweet that has been deleted, made private, or was tweeted by a user that has since made their account private, was suspended, or deleted their account is no longer accessible. 

The collection of this dataset was IRB-approved by USC. 

\subsection{Inquiries}
If you have technical questions about the data collection, please refer to the issues section in the GitHub repository or contact Emily Chen at \url{echen920@usc.edu}.

For any further questions about this dataset, please contact Dr. Emilio Ferrara at \url{emiliofe@usc.edu}.

\section{Acknowledgments}
The authors gratefully acknowledge support from DARPA (contract no. HR001121C0169). 
The authors also thank Kristina Lerman, Jon May, and our lab for helping to identify pertinent keywords to track in our ongoing data collection.

\bibliography{aaai23}

\begin{thebibliography}{43}
\providecommand{\natexlab}[1]{#1}

\bibitem[{oso(2022)}]{osome2022}
 2022.
\newblock Suspicious Twitter Activity around the Russian Invasion of Ukraine.
\newblock Technical report, Indiana University.

\bibitem[{Badawy et~al.(2019)Badawy, Addawood, Lerman, and
  Ferrara}]{badawy2019characterizing}
Badawy, A.; Addawood, A.; Lerman, K.; and Ferrara, E. 2019.
\newblock Characterizing the 2016 Russian IRA influence campaign.
\newblock \emph{Social Network Analysis and Mining}, 9: 1--11.

\bibitem[{Badawy, Ferrara, and Lerman(2018)}]{badawy2018analyzing}
Badawy, A.; Ferrara, E.; and Lerman, K. 2018.
\newblock Analyzing the digital traces of political manipulation: The 2016
  Russian interference Twitter campaign.
\newblock In \emph{2018 IEEE/ACM international conference on advances in social
  networks analysis and mining (ASONAM)}, 258--265. IEEE.

\bibitem[{Badawy, Lerman, and Ferrara(2019)}]{badawy2019falls}
Badawy, A.; Lerman, K.; and Ferrara, E. 2019.
\newblock Who falls for online political manipulation?
\newblock In \emph{Companion proceedings of the 2019 world wide web
  conference}, 162--168.

\bibitem[{Bigg(2023)}]{bigg_2023}
Bigg, M.~M. 2023.
\newblock Russia invaded Ukraine more than 10 months ago. here is one key
  development from every month of the war.

\bibitem[{Bloomberg(2022)}]{bloomberg2022visual}
Bloomberg. 2022.
\newblock A Visual Guide to the Russian Invasion of Ukraine.

\bibitem[{Broniatowski et~al.(2018)Broniatowski, Jamison, Qi, AlKulaib, Chen,
  Benton, Quinn, and Dredze}]{broniatowski2018weaponized}
Broniatowski, D.~A.; Jamison, A.~M.; Qi, S.; AlKulaib, L.; Chen, T.; Benton,
  A.; Quinn, S.~C.; and Dredze, M. 2018.
\newblock Weaponized health communication: Twitter bots and Russian trolls
  amplify the vaccine debate.
\newblock \emph{American journal of public health}, 108(10): 1378--1384.

\bibitem[{Bushwick(2022)}]{bushwick2022russia}
Bushwick, S. 2022.
\newblock Russia's information war is being waged on social media platforms.
\newblock \emph{Scientific American}.

\bibitem[{Cengel(2022)}]{cengel2022history}
Cengel, K. 2022.
\newblock The 20th-century history behind Russia's invasion of Ukraine.
\newblock \emph{Smithsonian.com}.

\bibitem[{Cheng and Anna(2022)}]{cheng_anna_2022}
Cheng, M.; and Anna, C. 2022.
\newblock WHO chief: Lack of help for Tigray Crisis due to Skin Color.

\bibitem[{Clinch(2022)}]{clinch2022russia}
Clinch, M. 2022.
\newblock How Russia invaded Ukraine in 2014. and how the markets tanked.
\newblock \emph{CNBC}.

\bibitem[{Cohen(2022)}]{cohen2022surge}
Cohen, R. 2022.
\newblock A surge of unifying moral outrage over Russia's war.
\newblock \emph{The New York Times}.

\bibitem[{Collett-White and Popeski(2014)}]{collett-white_2014crimeans}
Collett-White, M.; and Popeski, R. 2014.
\newblock Crimeans vote over 90 percent to quit Ukraine for Russia.
\newblock \emph{Reuters}.

\bibitem[{Dutt, Deb, and Ferrara(2019)}]{dutt2019senator}
Dutt, R.; Deb, A.; and Ferrara, E. 2019.
\newblock “Senator, We Sell Ads”: Analysis of the 2016 Russian Facebook Ads
  Campaign.
\newblock In \emph{Advances in Data Science: Third International Conference on
  Intelligent Information Technologies, ICIIT 2018, Chennai, India, December
  11--14, 2018, Proceedings 3}, 151--168. Springer.

\bibitem[{Ezzeddine et~al.(2022)Ezzeddine, Luceri, Ayoub, Sbeity, Nogora,
  Ferrara, and Giordano}]{ezzeddine2022characterizing}
Ezzeddine, F.; Luceri, L.; Ayoub, O.; Sbeity, I.; Nogora, G.; Ferrara, E.; and
  Giordano, S. 2022.
\newblock Characterizing and Detecting State-Sponsored Troll Activity on Social
  Media.

\bibitem[{Ferrara(2022)}]{ferrara2022twitter}
Ferrara, E. 2022.
\newblock Twitter Spam and False Accounts Prevalence, Detection and
  Characterization: A Survey.
\newblock \emph{First Monday}, (27).

\bibitem[{Ferrara et~al.(2020)Ferrara, Chang, Chen, Muric, and
  Patel}]{ferrara2020characterizing}
Ferrara, E.; Chang, H.; Chen, E.; Muric, G.; and Patel, J. 2020.
\newblock Characterizing social media manipulation in the 2020 US presidential
  election.
\newblock \emph{First Monday}.

\bibitem[{Funakoshi, Lawson, and Deka(2022)}]{funakoshi2022tracking}
Funakoshi, M.; Lawson, H.; and Deka, K. 2022.
\newblock Tracking sanctions against Russia.
\newblock \emph{Reuters}.

\bibitem[{Garner(2022)}]{garne2022war}
Garner, I. 2022.
\newblock How is the war going for Putin on social media? not great.
\newblock \emph{The Washington Post}.

\bibitem[{Hayda(2022)}]{hayda_2022}
Hayda, J. 2022.
\newblock Kyiv hosts a different kind of parade to celebrate Ukraine's
  Independence Day.

\bibitem[{Jiang et~al.(2020)Jiang, Chen, Yan, Lerman, and
  Ferrara}]{jiang2020political}
Jiang, J.; Chen, E.; Yan, S.; Lerman, K.; and Ferrara, E. 2020.
\newblock {Political Polarization Drives Online Conversations About COVID-19 in
  the United States}.
\newblock \emph{HBET}, 2: 200--211.

\bibitem[{John and Kesaieva(2022)}]{john_kesaieva_2022}
John, T.; and Kesaieva, Y. 2022.
\newblock Ukraine's Independence Day darkened by Deadly Missile Strike.

\bibitem[{Kahn(2022)}]{kahn2022viktor}
Kahn, J. 2022.
\newblock Who is Viktor Yanukovych? the ousted Ukrainian president that Putin
  hopes to put back in power.
\newblock \emph{Fortune}.

\bibitem[{La~Gatta et~al.(2023)La~Gatta, Wei, Luceri, Pierri, and
  Ferrara}]{la2023retrieving}
La~Gatta, V.; Wei, C.; Luceri, L.; Pierri, F.; and Ferrara, E. 2023.
\newblock Retrieving false claims on Twitter during the Russia-Ukraine
  conflict.
\newblock \emph{WWW’23 Companion Proceedings}.

\bibitem[{Luceri, Giordano, and Ferrara(2020)}]{luceri2020detecting}
Luceri, L.; Giordano, S.; and Ferrara, E. 2020.
\newblock Detecting troll behavior via inverse reinforcement learning: A case
  study of Russian trolls in the 2016 us election.
\newblock In \emph{Proceedings of the international AAAI conference on web and
  social media}, volume~14, 417--427.

\bibitem[{McGee and Princewill(2022)}]{mcgee2022world}
McGee, L.; and Princewill, N. 2022.
\newblock World leaders respond to Ukraine invasion, as fresh sanctions await
  Russia.
\newblock \emph{CNN}.

\bibitem[{{Media Bias/Fact Check}(2022)}]{mbfc}
{Media Bias/Fact Check}. 2022.
\newblock Media Bias/Fact Check.

\bibitem[{Mehrabi et~al.(2021)Mehrabi, Morstatter, Saxena, Lerman, and
  Galstyan}]{mehrabi2021survey}
Mehrabi, N.; Morstatter, F.; Saxena, N.; Lerman, K.; and Galstyan, A. 2021.
\newblock A survey on bias and fairness in machine learning.
\newblock \emph{ACM Computing Surveys (CSUR)}, 54(6): 1--35.

\bibitem[{Myers and Barry(2014)}]{myers2014putin}
Myers, S.~L.; and Barry, E. 2014.
\newblock Putin reclaims Crimea for Russia and bitterly denounces the west.
\newblock \emph{The New York Times}.

\bibitem[{Pereira and Reevell(2022)}]{pereira2022what}
Pereira, I.; and Reevell, P. 2022.
\newblock What to know about Ukrainian President Volodymyr Zelenskyy.
\newblock \emph{ABC News}.

\bibitem[{Pierri, Luceri, and Ferrara(2022)}]{pierri2022does}
Pierri, F.; Luceri, L.; and Ferrara, E. 2022.
\newblock How does Twitter account moderation work? Dynamics of account
  creation and suspension during major geopolitical events.
\newblock \emph{arXiv preprint arXiv:2209.07614}.

\bibitem[{Pierri et~al.(2022)Pierri, Luceri, Jindal, and
  Ferrara}]{pierri2022propaganda}
Pierri, F.; Luceri, L.; Jindal, N.; and Ferrara, E. 2022.
\newblock Propaganda and Misinformation on Facebook and Twitter during the
  Russian Invasion of Ukraine.
\newblock \emph{15th ACM Web Science Conference 2023}.

\bibitem[{Pitrelli(2022)}]{pitrelli2022anonymous}
Pitrelli, M.~B. 2022.
\newblock Global Hacking Group Anonymous launches 'Cyber War' against Russia.
\newblock \emph{CNBC}.

\bibitem[{Reuters(2022)}]{reuters2022timeline}
Reuters. 2022.
\newblock Timeline: The events leading up to Russia's invasion of Ukraine.
\newblock \emph{Reuters}.

\bibitem[{Schwirtz, Kramer, and Gladstone(2022)}]{schwirtz2022humanitarian}
Schwirtz, M.; Kramer, A.~E.; and Gladstone, R. 2022.
\newblock Humanitarian crisis worsens for Ukrainians trapped in Russia's
  onslaught.

\bibitem[{Schwirtz and Pérez-Peña(2022)}]{schwirtz2022first}
Schwirtz, M.; and Pérez-Peña, R. 2022.
\newblock First Ukraine City Falls as Russia strikes more civilian targets.
\newblock \emph{The New York Times}.

\bibitem[{Scott(2022)}]{scott2022war}
Scott, M. 2022.
\newblock As war in Ukraine evolves, so do disinformation tactics.
\newblock \emph{POLITICO}.

\bibitem[{Sharma et~al.(2021)Sharma, Zhang, Ferrara, and
  Liu}]{sharma2021identifying}
Sharma, K.; Zhang, Y.; Ferrara, E.; and Liu, Y. 2021.
\newblock Identifying Coordinated Accounts on Social Media through Hidden
  Influence and Group Behaviours.
\newblock In \emph{Proceedings of the 27th ACM SIGKDD Conference on Knowledge
  Discovery \& Data Mining}, 1441--1451.

\bibitem[{Stern et~al.(2022)Stern, Firozi, Paquette, Pannett, Francis, Dixon,
  Paúl, Knowles, and Kornfield}]{stern2022limited}
Stern, D.~L.; Firozi, P.; Paquette, D.; Pannett, R.; Francis, E.; Dixon, R.;
  Paúl, M.~L.; Knowles, H.; and Kornfield, M. 2022.
\newblock Limited cease-fire reached for civilian evacuations as Russian forces
  cut off Key Cities.
\newblock \emph{The Washington Post}.

\bibitem[{Stern, Suliman, and Taylor(2022)}]{stern2022russia}
Stern, D.~L.; Suliman, A.; and Taylor, A. 2022.
\newblock Russia not cooperating on proposed humanitarian corridor in Kherson,
  Ukrainian officials say.

\bibitem[{Sullivan(2022)}]{sullivan2022russia}
Sullivan, B. 2022.
\newblock Russia's at war with Ukraine. Here's how we got here.
\newblock \emph{NPR}.

\bibitem[{Walsh(2022)}]{walsh_2022}
Walsh, D. 2022.
\newblock After secret U.S. talks fail, a hidden war in Africa rapidly
  escalates.

\bibitem[{Westfall(2022)}]{westfall2022donetsk}
Westfall, S. 2022.
\newblock Why are Donetsk and Luhansk in Ukraine's Donbas region a flash point
  for Putin?
\newblock \emph{The Washington Post}.

\end{thebibliography}

\end{document}